\documentclass[aps,showpacs,amsmath,amssymb, nofootinbib]{revtex4}

\usepackage{graphicx}
\usepackage{bm}

\begin{document}

\date{\today}

\begin{center}
{\large {\bf Laue diffraction of  M\"{o}ssbauer and x-ray photons in strongly absorbing crystals}} \\
\vspace{0.5cm}
A.Ya.Dzyublik, V.Yu.Spivak\\
\vspace{0.5cm}
Institute for Nuclear Research of the Ukrainian Academy of
Sciences, avenue Nauki 47,
03680  Kiev, Ukraine
\end{center}
keywords: M\"{o}ssbauer spectroscopy, Laue diffraction, dynamical scattering theory,  $\gamma$-photon wave function, x-rays\\
\vspace{0.5cm}
\begin{center}
Abstract
\end{center}

 The dynamical scattering theory is developed for  the  Laue diffraction  of the M\"{o}ssbauer rays and x-rays, whose angular distribution is comparable
with the diffraction angular range. Both the Rayleigh and the resonant nuclear scattering are taken into account.
We consider typical case when incident radiation  first passes through an entrance slit and afterwards diffracts at the crystal planes within the Borrmann triangle.
  In calculations of  the wave function for $\gamma$-photons, refracted or diffracted in such strongly absorbing crystal, we apply the saddle-point method.
  The distribution of their  intensities over the basis of the Borrmann triangle is analyzed.
  In the spherical wave approximation of Kato, when  aperture of the incident beam much exceeds the diffraction interval, the derived formulae well correlate with the familiar equations of the diffraction theory of X-rays.

 \vspace{0.5cm}

\section{Introduction}
The diffraction of x-rays,
 synchrotron radiation, M\"{o}ssbauer rays and
neutrons is widely used for analysis of crystal structure. In this way such unique phenomena were
discovered as the pendell\"{o}sung
effect and the anomalous transmission  of $\gamma$-photons and
neutrons through a perfect crystal in the Laue (transmission)
geometry. In the x-ray optics the latter effect is frequently referred to as the Borrmann effect [1-3].
The explanation of these phenomena has been given by the dynamical
scattering theory [1-3].
In the two-wave case the incident plane x-ray wave generates  inside the crystal
 two couples of  waves, both of which are coherent superpositions of the transmitted and reflected waves.
One such couple has  nodes at the scattering atoms and is therefore anomalously weakly absorbed,
whereas another, having antinodes, is strongly absorbed.

The dynamical scattering theory has been extended to the case of elastic diffraction of M\"{o}ssbauer plane waves by  Afanas'ev and Kagan  \cite{Kagan}. They predicted that it can be realized complete suppression of $\gamma$-quanta by M\"{o}ssbauer nuclei in perfect crystals, that was confirmed in numerous experiments.

 Multiple scattering of x-ray photons by crystals is always
described by the Maxwell equations [1-3]. In the same
quasi-classical manner  Afanas'ev and Kagan \cite{Kagan} treated the
resonant scattering of M\"{o}ssbauer radiation by a
crystal. A quantum approach for the inelastic diffraction of $\gamma$-radiation in crystals, exposed to alternating external fields,
has been presented in \cite{Dz}.

In typical Laue-diffraction experiments the incident $\gamma$-quanta are first collimated by a slit, lying on the crystal
surface and being parallel to the reflecting planes (see Fig.1). And after that the  radiation flows within the angular region,
 which forms a so-called Borrmann
triangle (fan). The intensity distribution of the transmitted and reflected
beams over the
basis of the Borrmann triangle is analyzed with the aid of one
more slit, which is also parallel to the reflecting crystal
planes.

Standard plane-wave dynamical theory is not able to describe this situation. Therefore
Kato \cite{Kato1} considered the Laue diffraction of x-ray spherical waves, treating them like a
superposition of the classical  plane waves, which
spread over the angle $\theta$ about the Bragg angle $\theta_{\textrm{\scriptsize {B}}}$.
Every such plane component independently of each other are scattered by  atoms of the crystal, forming
the refracted and diffracted wave packets, represented by  by the integrals
over the angle $\theta$. I such a spherical-wave approximation it was silently believed that the dispersion $\sigma$ of the incident rays over $\theta$ much exceeds the characteristic angular interval $|\Delta\vartheta|$, where the diffraction proceeds.
Kato found exact solution of the integral over $\theta$ in terms of the Bessel function.
Besides,
these integrals can be estimated over $\theta$ with the aid of the stationary-phase
approximation if the crystal  thickness $D$ to be much larger than
the pendell\"{o}sung length $\Lambda_{\textrm{\scriptsize {L}}}$ \cite{Authier}. This method, however, can be used only if the large parameter of the task $D/\Lambda_{\textrm{\scriptsize {L}}}$ is a real number.
In other words, it can be applied only to weakly absorbing crystals.

\begin{figure}
\includegraphics[width=9cm]{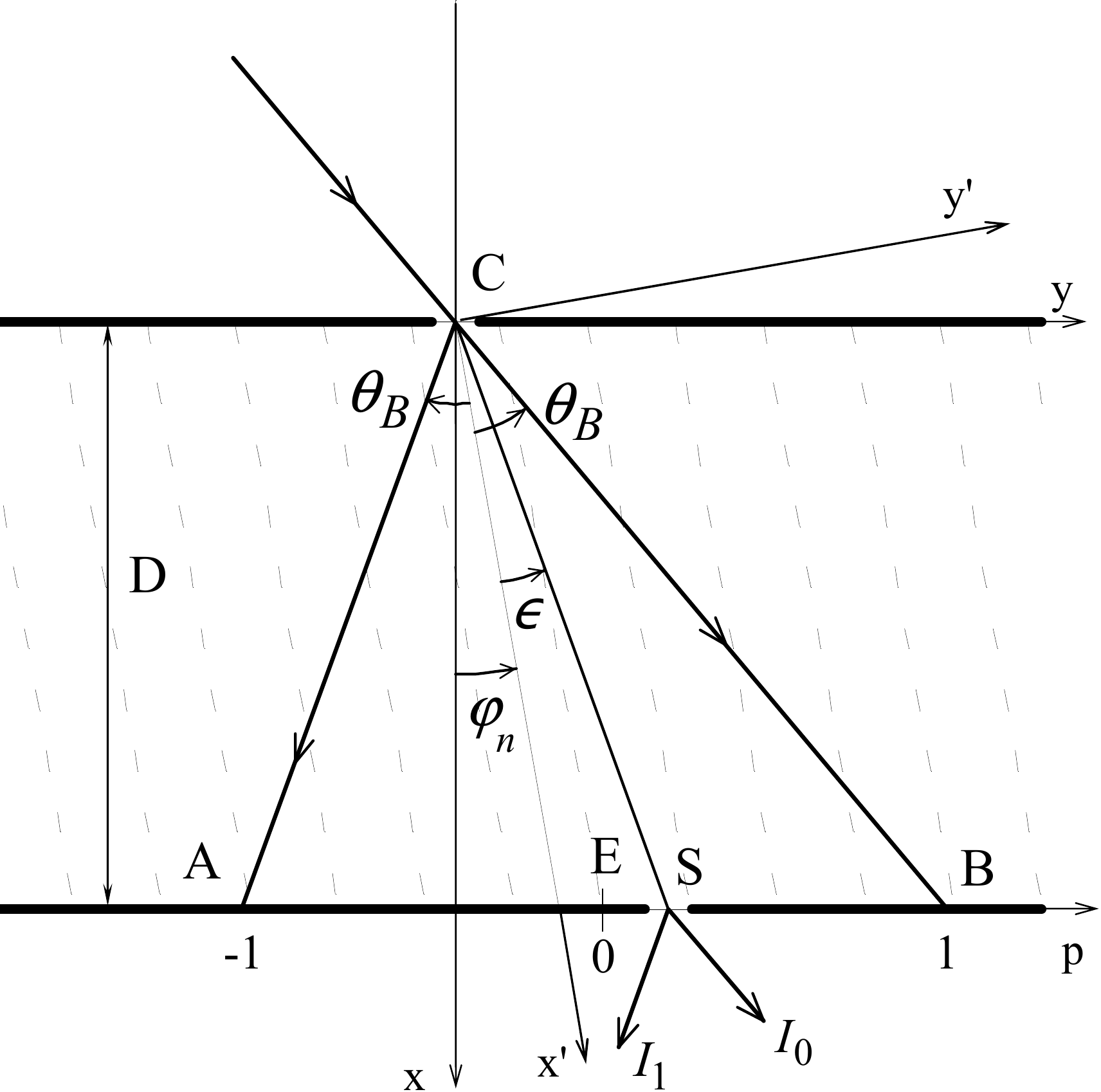}
\caption{Scheme of the Laue diffraction of the collimated $\gamma$-rays, whose
 flow is concentrated mainly inside  the Borrmann triangle ABC.
 The points A and B are marked by the reduced coordinates $p=- 1$ and $p=1$, respectively;
  the middle point E of the line segment AB by $p=0.$ The collimating and scanning slits are labeled by C and S, respectively.
The reflecting planes are drawn by the dashed lines.}
\end{figure}

At the same time,  for the M\"{o}ssbauer
diffraction, when the resonant scattering amplitude of
$\gamma$-quanta is already a complex number, the stationary-phase
method is not applicable. In this case such integrals should be
estimated in more general  saddle-point approach. Previously we
did this in the spherical-wave approximation, confining ourselves with
 the symmetric Laue diffraction of both M\"{o}ssbauer radiation \cite{DS} and neutrons \cite{DSM}.

Now we analyze much more general case of the M\"{o}ssbauer diffraction in arbitrary Laue geometry, when the dispersion $\sigma$ may be of the order of the diffraction range $|\Delta\vartheta|$.
In our calculations we describe photons by  the vector wave function ${\boldsymbol \Psi}({\bf r},t)$ suggested by Bialynicki-Birula
\cite{P1}. It  is represented by the wave packet, composed by
 the   plane waves ${\boldsymbol \chi}_{{\boldsymbol\kappa}\lambda}({\bf r}) e^{-i\omega t}$, where
\begin{equation}\label{eq:pwf}
{\boldsymbol \chi}_{{\boldsymbol\kappa}\lambda}({\bf r})={\bf
e }_{\lambda}({\boldsymbol\kappa}) \sqrt{\hbar \omega}e^{i{\boldsymbol\kappa}
{\bf r}}
\end{equation}
with  the wave vector ${\boldsymbol\kappa}$, helicity  $\lambda=\pm 1$ and frequency $\omega=c\kappa$.
The modulus squared of ${\boldsymbol \Psi}({\bf r},t)$ is now interpreted as a
probability density of detecting the photon's mean energy  at the given
position ${\bf r}$ in the moment $t$. Application of such a wave function allows us to employ quantum scattering theory \cite{Goldberger, Sitenko} for analysis of the diffraction of $\gamma$-quanta.

\section{Scattering amplitudes}
We introduce the  right-hand coordinate frame $x,\;y,\;z$  with the origin on the entrance
 surface  in the middle of the collimating slit. The axis $x$ is directed inside the crystal perpendicularly to its
 surface and the axis $z$  along the slit (see Fig.1).
 One introduces also the frame $x',\;y',\;z'$  with the axis $x'$  parallel to  the reflecting crystal planes and axis $z'$ coinciding with $z$. It is obtained from $x,\;y,\;z$ by  rotation through the angle $\varphi_n$ around the axis $z$. Here $\varphi_n<0$ for clockwise rotation  and $\varphi_n>0$ otherwise. The angle between the photon wave vector ${\boldsymbol\kappa}$  and the axis $x$ is $\varphi$, the  incidence angle on the reflecting planes is $\theta$, the Bragg angle $\theta_{\textrm{\scriptsize {B}}}.$

  In addition, we introduce the angles $\varphi_0$  and $\varphi_1$  between the axis $x$ and the sides of the Borrmann triangle,  $\varphi_0=\theta_{{\scriptsize {B}}}+\varphi_n >0$ and
 $\varphi_1=-\theta_{\textrm{\scriptsize {B}}}+\varphi_n<0 $.

 Let the divergent beam of $\gamma$-quanta move in the plane $x,~y$ perpendicularly to the slit and be spread over the angle $\theta$. Then the components of ${\boldsymbol\kappa}(\theta)$, written in cylindrical coordinates, are
   \begin{equation}\label{}
{\boldsymbol \kappa }(\theta)=\{\kappa\cos\varphi,\; \kappa\sin\varphi,\; 0\},
  \end{equation}
where $\varphi=\varphi_n+\theta$.

The wave function of the incident photon  may be written as
\begin{equation}\label{eq:WP1}
{\boldsymbol\Psi}_{s}({\bf r},t)_{{\textrm{in}}}=\int_0^{\infty} G_{e}(\omega){\boldsymbol\Psi}_{\omega,s}({\bf r})_{{\textrm{in}}}e^{-i\omega t}d\omega,
\end{equation}
where $G_{e}(\omega)$ specifies the frequency  distribution of the
incident photons, the function
\begin{equation}\label{eq:Psi}
{\boldsymbol\Psi}_{\omega,s}({\bf r})_{{\textrm{in}}}=\int_{-\pi}^{\pi}
G_a(\theta){\boldsymbol\chi}_{{\boldsymbol \kappa}(\theta) s}({\bf r}) d\theta
\end{equation}
  describes photons with fixed  frequency $\omega$ and polarization ${\bf e}_s$.
  The polarization vectors ${\bf e}_s({\boldsymbol \kappa})$ are perpendicular to the wave vector ${\boldsymbol \kappa}={\boldsymbol \kappa}(\theta)$ and therefore also depend on the angle $\theta$. However, one can neglect this dependence since  the dispersion $\sigma <<1$.

 We approximate  the angular distribution by the Gaussian function, concentrated at the angle $\theta_0$
  close to  the Bragg angle $\theta_{\textrm{\scriptsize {B}}}$:
\begin{equation}\label{eq:Gauss}
G_a(\theta)=\frac{1}{(\sqrt{2\pi}\sigma)^{1/2}}\exp\left\{-\frac{(\theta_{0}-\theta)^2}{4\sigma^2}\right\},
\end{equation}
where
\begin{equation}\label{}
\sigma^2=\langle (\theta_0-\theta)^2 \rangle
\end{equation}
 denotes the mean-square angular distribution of the beam.
Usually $\sigma<<1$, that enables us to spread the integration limits over $\theta$ from ${-\infty}$ to $\infty$.
In the spherical-wave approximation of Kato this distribution is replaced by unity.

In general case polarizations of $\gamma$-quanta are mixed during the resonant scattering by nuclei.
Below  we only consider scattering by unpolarized   M\"{o}ssbauer nuclei with unsplit sublevels, labeled by the magnetic quantum number.
In this case the mixing is avoided if $\gamma$-quanta have either    $\pi$-polarization ${\bf e}_{s=1}$, lying in the scattering plane or
  $\sigma$-polarization ${\bf e}_{s=2}$, being  perpendicular to this plane \cite{Belyakov}.

The frequency distribution  for the phononless emission line is determined by  the function
\begin{equation}
 G_e(\omega)=\left(\frac{\Gamma_e}{2\pi\hbar}\right)^{1/2} \frac{e^{i\omega
t_0}}{\omega-\omega_e+i\Gamma_{e}/2\hbar},
\end{equation}
where
 $\Gamma_{e}$ and $\hbar\omega_e$ are respectively the width
 and energy of the excited  level of the emitting nucleus.
Time dependence  of the incident $\gamma$-quantum  at the entrance surface $(x=0)$ is described by the exponential:
\begin{equation}\label{eq:exp}
|{\boldsymbol \Psi}^{(s)}_{\textrm{\scriptsize {in}}}(0,t)|^2
\sim \exp[-\Gamma_e(t-t_0)/\hbar]\theta(t-t_0),
\end{equation}
where
\begin{equation}\label{eq:7}
 \theta(x)= \left\{\begin{array}{lll} 1,&\;&\;x>0,\\ 0,&\;&\;x<0.
 \end{array}\right.
\end{equation}
is the Heaviside step function. According to (\ref{eq:exp}) $t_0$ means the   moment of the  photon's arrival  to the crystal.

The coherent scattering amplitude of $\gamma$-quanta by the $j$th nucleus with unsplit sublevels
is given by \cite{Belyakov}
\begin{equation}\label{eq:11l}
f_{\textrm{\scriptsize
{coh}}}({\boldsymbol \kappa},{\bf e}_s;{\boldsymbol \kappa}',{\bf e}'_s)^{N}_{j}= -c_0\left(\frac{%
2I_e+1}{2I_{g}+1}\right)
\frac{1}{4\kappa}
 \frac{\Gamma_{\gamma}e^{-W_{j}({\boldsymbol \kappa})-W_{j}({\boldsymbol \kappa}')}}{\hbar(\omega-\omega'_0)
+i\Gamma/2}P_s^N,
\end{equation}
 where $c_0$ is the relative concentration of  the M\"{o}ssbauer
isotope,  $I_e$ and $I_g$ are the nuclear spins in the excited and
ground states, $e^{-W_j({\bf k})}$~is the Lamb-M\"{o}ssbauer factor, $P_s^N$ is the nuclear polarization factor,
$\Gamma$ and $\Gamma_{\gamma}$ are respectively  the total  and
radiative widths of the resonant nuclear level with the energy
$\hbar\omega'_0$.
In the case of M1 transitions  $P_s^N=1$ for the  $\pi$-polarization, and
$P_s^N={\bf e}\cdot {\bf e}'$ for the  $\sigma$-polarization \cite{Belyakov}.

The coherent Rayleigh scattering amplitude by the $j$th atom is
determined by the expression
\begin{equation}\label{}
f_{\textrm{\scriptsize {coh}}}({\boldsymbol \kappa},{\bf e}_s;{\boldsymbol \kappa}',{\bf
e}'_s)_{j}^{R}=  e^{-W_{j}({\bf Q})}r_0 F_e^{(j)}({\bf
Q})({\bf e}_s\cdot{\bf e}'^{*}_s)+ (\kappa/4\pi)\sigma^{(j)}_{e},
\end{equation}
where the form-factor of the $j$th atom
\begin{equation}\label{}
F_e^{(j)}({\bf Q})=\int \varrho^{(j)}_e({\bf r})e^{i{\bf Qr}}d{\bf r},
\end{equation}
${\bf Q}={\boldsymbol \kappa}-{\boldsymbol \kappa}'$ is the scattering vector,
 $\varrho^{(j)}_e({\bf r})$ is the  density of  atomic electrons,
$r_0=e^2/mc^2$ denotes the classical radius of the electron,
$\sigma^{(j)}_{e}$ is the  absorption cross section of $\gamma$-quanta by electrons of the $j$th atom.

 The coherent scattering amplitude of $\gamma$-quanta by
 elementary cell of the crystal
\begin{equation}\label{}
F_{\textrm{\scriptsize{coh}}}
({\boldsymbol \kappa},{\bf e};{\boldsymbol \kappa}',{\bf e}')=
F_{\textrm{\scriptsize
{coh}}}({\boldsymbol \kappa},{\bf e};{\boldsymbol \kappa}',{\bf e}')^{N}
+F_{\textrm{\scriptsize
{coh}}}({\boldsymbol \kappa},{\bf e};{\boldsymbol \kappa}'
,{\bf e}')^{R},
\end{equation}
where
\begin{equation}\label{}
F_{\textrm{\scriptsize {coh}}}
({\boldsymbol \kappa},{\bf e};{\boldsymbol \kappa}',{\bf
e}')^{N(R)}=
\sum_{j}e^{i{\bf Q}{\boldsymbol\rho}_{j}} f_{\textrm{\scriptsize {coh}}}
({\boldsymbol \kappa},{\bf e};{\boldsymbol \kappa}',{\bf e}')_{j}^{N(R)}
\end{equation}
are the
nuclear and Rayleigh coherent scattering amplitudes, depending on
 the radius vector ${\boldsymbol \rho}_{j}$ for the equilibrium
position of the $j$th atom in the elementary cell.

\section{Dynamical scattering theory}
 Every  plane component ${\boldsymbol\chi}_{{{\boldsymbol\kappa}(\theta),s}}({\bf r})$ of the incident wave packet (\ref{eq:Psi})
is scattered independently, generating the  wave
${\boldsymbol\psi}_{{{\boldsymbol\kappa}(\theta),s}}({\bf r})$  \cite{Goldberger}. As a consequence,  the complete  wave function of the photon will be
\begin{equation}\label{eq:120}
{\boldsymbol \Psi}_{s}({\bf r},t)=\int_{-\infty}^{\infty}d\omega G_e(\omega)
\int_{-\infty}^{\infty}d\theta G_a(\theta)
{\boldsymbol \psi}_{{\boldsymbol \kappa} s}({\bf r})e^{-i\omega t}
\end{equation}
with the same distributions $G_e(\omega)$ and $ G_a(\theta)$ as the incident wave packet (\ref{eq:WP1}), (\ref{eq:Psi}).

In the two-wave diffraction  the wave  ${\boldsymbol \psi}_{{\boldsymbol \kappa }(\theta)}({\bf r})$ inside the crystal  as $0<x<D$, where $D$ is the crystal thickness, consists of the refracted wave with the wave vector
 ${\bf k}(\theta)$ and the diffracted one with the wave vector ${\bf k}_{1}(\theta)={\bf k}(\theta)+{\bf h}$,  where ${\bf h}$ denotes a reciprocal lattice vector. The components of the vectors ${\bf k}(\theta)$ and ${\boldsymbol\kappa}(\theta)$
along the entrance surface $x=0$ coincide and therefore
\begin{equation}\label{eq:}
{\bf k}_{\nu}(\theta)={\boldsymbol\kappa}_{\nu}(\theta) +\delta(\theta){\bf n},\qquad
{\boldsymbol\kappa}_{1}(\theta)={\boldsymbol\kappa}_{0}(\theta)+{\bf h},
\end{equation}
where ${\bf n}$ is the unit vector along the axis $x$.

As a consequence,  the wave function $ {\boldsymbol\Psi}_{\omega s }({\bf r})$ inside the crystal transforms to
\begin{equation}\label{eq:Psi1}
{\boldsymbol\Psi}_{\omega s}({\bf r})=
\sum_{\nu=0,1}{\boldsymbol\Psi}_{\omega s}^{(\nu)}({\bf r}),
\end{equation}
where
\begin{equation}\label{eq:psi10}
{\boldsymbol\Psi}_{\omega s}^{(\nu)}({\bf r})=\int_{-\infty}^{\infty} d\theta G_a(\theta)
{\boldsymbol\psi}_{{\boldsymbol \kappa}_{\nu}(\theta),s}({\bf r})
\end{equation}
with ${\boldsymbol\psi}_{{\boldsymbol \kappa}_{\nu}(\theta),s}({\bf r})$ given by
\begin{equation}\label{eq:66}
 {\boldsymbol\psi}_{{\boldsymbol \kappa}_{\nu}(\theta),s}({\bf r})={\bf e}_{\nu s}\sqrt{\hbar\omega}
 \sum_{\iota=1,2}C_{\nu s }^{(\iota)}(\theta)
e^{i{\boldsymbol\kappa}_{\nu}(\theta){\bf r}+i\delta_{\iota s  }(\theta)x}.
\end{equation}

For the two-wave case  the amplitudes $C$ and the wave vectors ${\bf k}$ are determined by   equations
\cite{Kagan}
\begin{eqnarray}\label{eq:fe}
 (k^2(\theta)/\kappa^2(\theta)-1)C_{0}=g_{00}C_{0}+g^{(s)}_{01}C_{1},\\
(k^2_1(\theta)/\kappa^2(\theta)-1)C_{1}=g^{(s)}_{10}C_{0}+g_{11}C_{1}.\nonumber
\end{eqnarray}
The scattering   matrix $g^{(s)}_{\mu\nu}$ is defined by the expression
\begin{equation}\label{eq:3}
g^{(s)}_{\mu\nu}=\frac{4\pi }{\kappa^2
v_0}F({{\boldsymbol\kappa}_{\nu},{\bf e}_{\nu s};{\boldsymbol\kappa}_{\mu}},{\bf e}_{\mu s}),\qquad\qquad
\mu,\nu=0,1,
\end{equation}
where $v_0$  stands for the volume of the elementary cell.

The system of two  equations (\ref{eq:fe}) has the
following solution  \cite{Kagan}:
\begin{eqnarray}\label{eq:23}
\delta^{(s)}_{1,2}=\kappa\varepsilon_{0s}^{(1,2)}/\gamma_0,
\end{eqnarray}
where
\begin{eqnarray}\label{eq:23a}
\varepsilon_{0s}^{(1,2)}=\frac{1}{4}\left[g_{00}+
\beta g_{11}-\beta\alpha \right]
\pm \frac{1}{4}\left\{\left[g_{00}+\beta g_{11}
-\beta\alpha\right]^2
+4\beta\left[g_{00}\alpha
-(g_{00}g_{11}-g^{(s)}_{01}g^{(s)}_{10})\right]\right\}^{1/2},
\end{eqnarray}
with
\begin{equation}
 \beta=\gamma_0/\gamma_1,\qquad
 \alpha=\frac{2{\boldsymbol\kappa}(\theta){\bf h}+{\bf h}^2}{\kappa^2},
\qquad \gamma_{\nu}=\cos\varphi_{\nu}.
\end{equation}
The angle $\alpha$ defines   deviation from the exact Bragg condition  $\kappa_1=\kappa$.
For photons with fixed frequency \cite{Zach}
\begin{equation}\label{eq:Z}
\alpha=2\sin2\theta_B\Delta\theta,
\end{equation}
where
\begin{equation}\label{eq:1d}
 \Delta\theta=\theta_{\textrm{\scriptsize {B}}}-\theta.
\end{equation}

It is most convenient to  express $\varepsilon_{0}^{(1,2)}$ in terms of new deviation parameter
\begin{equation}\label{eq:4}
\eta=\frac{1}{2}\left(\frac{\beta}{g^{(s)}_{01}g^{(s)}_{10}}\right)^{1/2}(\alpha-\alpha_0),
 \end{equation}
 where the angular shift
 \begin{equation}\label{}
 \alpha_0=g_{11}  -g_{00}/\beta.
 \end{equation}
 Notice that $\alpha_0$ diminishes in the case of symmetric diffraction, $\beta=1$, if $g_{11}=g_{00}$.

Making use of the above definitions we transform the parameter  $\delta_{\iota}(\eta)$, defined by Eqs.(\ref{eq:23}), (\ref{eq:23a}), to
  \begin{equation}\label{eq:delta}
\delta^{(s)}_{\iota}(\eta)=\frac{\kappa g_{00}}{2\gamma_0}-\frac{\pi}{\Lambda_L}\left[\eta+(-1)^{\iota+1}\sqrt{1+\eta^2}\right],
\end{equation}
where
\begin{equation}\label{}
\Lambda_L=\frac{2\pi \gamma_0}{\kappa \sqrt{ g^{(s)}_{01}g^{(s)}_{10}\beta}}
\end{equation}
means the Pendell\"{o}sung distance in the case of weakly
absorbing crystals  (see, e.g.,
\cite{Authier}).

For the Laue diffraction ($\beta>0$) the amplitudes of the waves satisfy the following boundary conditions at $x=0$:
\begin{eqnarray}\label{}
\sum_{\iota=1,2}C_{0s}^{(\iota)}(\theta)=1, \qquad
 \sum_{\iota=1,2}C_{1s}^{(\iota)}(\theta)=0.
\end{eqnarray}
Being  expressed in terms of $\eta$, these amplitudes  take the form
\begin{eqnarray}\label{eq:C}
C_{0s}^{(\iota)}(\eta) &=\frac{1}{2}\left[1+(-1)^{\iota}\frac{\eta}{\sqrt{1+\eta^2}}\right], \qquad
C_{1s}^{(\iota)}(\eta) &=\frac{(-1)^{\iota}}{2}\left(\frac{g^{(s)}_{10}}{g^{(s)}_{01}}\right)^{\frac{1}{2}}\sqrt{\frac{\beta}{1+\eta^2}}.
\end{eqnarray}

 Combining Eqs. (\ref{eq:Z}) and (\ref{eq:4}), one finds the relation between our  departure parameters:
\begin{equation}\label{eq:333}
\Delta\vartheta\eta=\theta'_{\textrm{B}}-\theta,
 \end{equation}
where
\begin{equation}\label{}
\Delta\vartheta=\frac{1}{\sin 2\theta_{\textrm{B}}}
\sqrt{\frac{g^{(s)}_{01}g^{(s)}_{10}}{\beta}},\qquad  \theta'_{\textrm{B}}=\theta_{\textrm{B}}+\Delta \theta_{\textrm{B}},\qquad
\Delta\theta_{\textrm{B}}=-\frac{\alpha_0}{2\sin 2\theta_{\textrm{\scriptsize {B}}}}.
\end{equation}
From (\ref{eq:333}) we see that $|\Delta\vartheta|$ implies a  characteristic diffraction interval, where $|\eta|\leq 1,$
while $\theta'_{\textrm{B}}$ stands for the corrected Bragg angle corresponding to $\eta=0$.
The above formulas can be also rewritten as
\begin{equation}\label{eq:rev}
 \Delta\theta = \frac{\alpha_0}{2\sin 2\theta_B}+\Delta\vartheta\eta.
\end{equation}

 Using the parameters introduced above one can transform the expression (\ref{eq:psi10}) for the wave function to
\begin{equation}\label{eq:psi0}
{\boldsymbol\Psi}_{\omega s}^{(\nu)}({\bf r})=-\sqrt{\Delta\vartheta}\int_{-\infty}^{\infty}  d\eta{\cal G}_a(\eta)
{\boldsymbol\psi}_{{\boldsymbol \kappa}_{\nu}(\theta),s}({\bf r}).
\end{equation}
Now
the angular distribution versus $\eta$ is given by
\begin{equation}\label{eq:G}
{\cal G}_a(\eta)=  \frac{1}{(2\pi \overline{\eta^2})^{1/4}}
 \exp\left\{-\frac{(\eta-\Delta\eta)^2}{4\overline{\eta^2}}\right\},
\end{equation}
where the parameter
\begin{equation}\label{}
\Delta\eta=\frac{\theta'_{\textrm{B}}-\theta_0}{\Delta\vartheta}
\end{equation}
 characterizes a mismatch of the beam orientation and the Bragg resonance, and
\begin{equation}\label{}
\overline{\eta^2}=\left(\sigma/\Delta\vartheta\right)^2
\end{equation}
specifies a squared width of the distribution.

The intensity distribution over the basis of the Borrmann
triangle is usually analyzed with the aid of the scanning slit,
located on the rear surface and  directed  along the axis z. Let this slit cross the axis $y$ in the point $y_S$, while
   the
midpoint E on the  basis AB of the Borrmann triangle have the coordinate $y_0$. It is convenient to introduce the reduced coordinate $p$
of the scanning slit determined by
\begin{equation}\label{eq:p11}
p=\frac{\Delta y_S}{L},
\end{equation}
where $2L$ is the length of the line segment AB and $\Delta y_S=y_S-y_0$ stands for  the shift if the scanning slit with respect to the center of the Borrmann triangle basis. The analogous definition of this reduced coordinate was given by Authier \cite{Authier}.
The definition (\ref{eq:p11}) is
equivalent to
\begin{equation}\label{eq:p}
p=2\frac{\Delta y_S/D}{\tan\varphi_0-\tan\varphi_1},
\end{equation}
which  in the case of symmetric diffraction, as $\beta=1$,  reduces to  the
definition of $p$, given in Refs.~\cite{Batterman, Kato1}:
\begin{equation}\label{}
p=\frac{\tan\epsilon}{\tan\theta_{\textrm{\scriptsize {B}}}},
\end{equation}
where  $\epsilon$ is the angle between the reflecting planes and
the direct line CS,  connecting  the entrance and exit  slits
(see Fig.~1).

Let us expand now the exponent of the $e^{i{\boldsymbol\kappa}(\theta){\bf r}}$ in $\Delta\theta$:
\begin{equation}\label{eq:e2}
e^{i{\boldsymbol\kappa}_{\nu}(\theta){\bf r}}\approx  \exp\left\{i\kappa\sin\varphi_0 D\left[1-\frac{y_S/D}{\tan\varphi_0}
   \right]\Delta\theta   \right\}e^{i{\boldsymbol\kappa}_{\nu}{\bf r}}.
\end{equation}
Here
\begin{equation}\label{eq:p5}
\frac{y_S/D}{\tan\varphi_0}= \frac{1}{\tan \varphi_0}\left[\frac{\tan\varphi_0+\tan\varphi_1}{2}+
\frac{\tan\varphi_0-\tan\varphi_1}{2} p  \right].
\end{equation}

From the equalities
\begin{equation}
\varphi_0+\varphi_1=2\varphi_n, \qquad \varphi_0-\varphi_1=2\theta_B
\end{equation}
it follows that
\begin{eqnarray}
\tan\varphi_0+\tan\varphi_1=\frac{\sin 2\varphi_n}{\gamma_0\gamma_1}
\end{eqnarray}
and
\begin{eqnarray}
\tan\varphi_0-\tan\varphi_1=\frac{\sin 2\theta_B}{\gamma_0\gamma_1}.
\end{eqnarray}

This allows us to transform (\ref{eq:p5}) as
\begin{equation}\label{eq:xz}
\frac{y_S/D}{\tan\varphi_0}=\frac{1}{2\tan\varphi_0}\left[\frac{\sin 2\varphi_n+p\sin 2\theta_B }{\gamma_0\gamma_1}\right].
\end{equation}
By making use of the relation
\begin{equation}\label{}
\sin 2\varphi_n+\sin 2\theta_B =2\sin\varphi_0\gamma_1
\end{equation}
one gets
\begin{equation}\label{}
\frac{y_S/D}{\tan\varphi_0}= 1+ \frac{\sin\theta_B\cos\theta_B}{\sin\varphi_0\gamma_1}(p-1).
\end{equation}

With this formula and  Eq.~(\ref{eq:rev}) we transform  Eq.~(\ref{eq:e2}) to
\begin{eqnarray}\label{eq:pw}
\exp\{{i{\boldsymbol\kappa}_{\nu}(\theta){\bf r}}\} =
\exp\left\{i\frac{\kappa D}{4\gamma_1}(1-p)\alpha_0\right\}\nonumber
\exp\left\{i\frac{\pi D}{\Lambda_L}(1-p)\eta\right\}e^{i{\boldsymbol\kappa}_{\nu}{\bf r}}.\qquad\qquad
\end{eqnarray}

Taking also into account Eq.~(\ref{eq:delta}), we are led  to the following expression for the plane waves at the exit slit, i.e., at ${\bf r}\approx
\{D,\;y_S,\;0\}$:
\begin{eqnarray}\label{eq:psi8}
\exp\{{i{\boldsymbol\kappa}_{\nu}(\theta){\bf r}+i\delta_{\iota}(\theta)D}\} = \Phi(p;\omega)
\exp\left\{-i\frac{\pi D}{\Lambda_L}\left[p\eta+(-1)^{\iota+1}\sqrt{1+\eta^2}\right] \right\}e^{i{\boldsymbol\kappa}_{\nu}{\bf r}},\nonumber
\end{eqnarray}
where we used the abbreviation
\begin{equation}
\Phi(p;\omega)=\exp\left\{i\frac{\kappa D}{4}\left[\frac{g_{00}} {\gamma_0}
+\frac{g_{11}}{\gamma_1}
 +p\left(\frac{g_{00}} {\gamma_0}-\frac{g_{11}} {\gamma_1} \right)\right]\right\}.
 \end{equation}

Inserting (\ref{eq:psi8}) into  (\ref{eq:psi0}) we represent the photon wave functions as
\begin{eqnarray}\label{eq:10}
{\boldsymbol\Psi}_{\omega,s}^{(\nu)}({\bf r})=-\sqrt{\hbar\omega}{\bf e}_s\Phi(p;\omega)\sqrt{\Delta\vartheta}
\sum_{\iota=1,2}
{\cal I}_{\nu s}^{(\iota)} e^{i{\boldsymbol\kappa}_{\nu}{\bf
r}},
\end{eqnarray}
 where
 ${\cal I}_{\nu s}^{(\iota)}$ denotes the integral
\begin{equation}\label{eq:int}
{\cal I}_{\nu s}^{(\iota)}=\int_{C}d\eta{\cal G}_a(\eta)
C_{\nu s}^{(\iota)}(\eta) \exp\{{\cal N}_s
{\cal S}_{\iota s}(\eta)\},
\end{equation}
the integration path $C$ on the complex plane $\eta=\eta_r+i\eta_i$ is along the  line,   defined by
  Im~$[\alpha_0+2(g^{(s)}_{01}g^{(s)}_{10}/\beta)^{1/2}\eta]=0,$
while
  \begin{equation}
{\cal N}_s=\frac{\pi D}{|\Lambda_L|}, \qquad
{\cal S}_{\iota s}(\eta)=-i\left(|\Lambda_L|/\Lambda_L\right) \left[p\eta+
(-1)^{\iota+1}\sqrt{1+\eta^2}\right].
\end{equation}

We assume that the crystal thickness $D>>|\Lambda_L|/\pi$ and the dispersion
\begin{equation}\label{}
\sigma>>2\sqrt{\frac{|\Lambda_L|}{\pi D}  }|\Delta\vartheta|.
\end{equation}
 This allows us to
estimate the integral (\ref{eq:int}) by the saddle-point method
(for details see \cite{DSM1}).
The saddle points determined by the equation ${\cal S}'_{\iota}(\eta)=0$
 are
\begin{equation}\label{}
\eta^{(\iota)}_0=(-1)^{\iota}\eta_0,\qquad\qquad   \eta_0=\frac{p}{\sqrt{1-p^2}}.
\end{equation}

The amplitudes $C_{\nu}^{(\iota)} =C_{\nu}^{(\iota)}(\eta^{\iota}_0)$ in the saddle points take the form
\begin{eqnarray}\label{eq:90}
C_{0}^{(\iota)}=\frac{1}{2}\left( 1+p\right), \qquad\qquad
C_{1}^{(\iota)}=\frac{(-1)^{\iota}}{2}\left(\frac{g_{10}}{g_{01}}\beta\right)^{\frac{1}{2}}\sqrt{1-p^2}.
\end{eqnarray}

In general,  the angular distribution (\ref{eq:G}) in the saddle points is represented by different factors
\begin{equation}\label{}
{\cal G}_1\equiv {\cal G}_a(\eta_0^{(1)})=\frac{1}{(2\pi \overline{\eta^2})^{1/4}}\exp\left\{-\frac{(\eta_0+\Delta\eta)^2}{4\overline{\eta^2}} \right\},\qquad
 {\cal G}_2\equiv {\cal G}_a(\eta_0^{(2)})=\frac{1}{(2\pi \overline{\eta^2})^{1/4}}\exp\left\{-\frac{(\eta_0-\Delta\eta)^2}{4\overline{\eta^2}} \right\},
\end{equation}
which only coincide in the case of exact Bragg resonance.

Employing standard formulas of the saddle-point method \cite{Math},
 one gets  the wave function of photons in any point ${\bf r}\approx {\bf r}_S $ close to the
scanning slit. For the refracted $\gamma$-quanta with fixed frequency $\omega$ the wave function is
\begin{eqnarray}\label{eq:F1}
{\boldsymbol\Psi}_{\omega s}^{(0)}({\bf r})= \frac{1}{2}\frac{{\bf
A}_{0s}(p)\Phi(p)}{(1-p^2)^{1/4}} \left(
\frac{1+p}{1-p}\right)^{\frac{1}{2}}
\sqrt{\frac{2\Lambda_L}{D}}\left[e^{i\zeta(p)}+e^{-i\zeta(p)}
\right]
 e^{i{\boldsymbol \kappa}_0{\bf r}}
\end{eqnarray}
and  for the diffracted  those
\begin{eqnarray}\label{eq:F2}
{\boldsymbol\Psi}_{\omega s}^{(1)}({\bf r})= \frac{1}{2}\frac{{\bf
A}_{1s}(p)\Phi(p)}{(1-p^2)^{1/4}}
\sqrt{\frac{2\Lambda_L}{D}} \left[e^{i\zeta(p)}-e^{-i\zeta(p)}\right]
 e^{i{\boldsymbol \kappa}_1{\bf r}},\nonumber
\end{eqnarray}
where the function $\zeta(p)$, depending on the phases $\phi_{\nu}=$arg$~{\cal G}_{\nu}$, is given by
\begin{equation}\label{eq:z}
 \zeta(p)=\frac{\pi D}{\Lambda_L}\sqrt{1-p^2}+\frac{\phi_2-\phi_1}{2}+ib+\frac{\pi}{4}
\end{equation}
with
\begin{equation}\label{}
b=\frac{1}{2}\ln\left|\frac{{\cal G}_1}{{\cal G}_2}  \right|,
\end{equation}
whereas the amplitudes ${\bf{ A}}_{\nu s}(p)$ are
\begin{eqnarray}\label{}
{\bf{ A}}_{0s}(p)=-{\bf e}_{0s}\sqrt{\hbar\omega}({\cal G}_1{\cal G}_2)^{1/2}\sqrt{\Delta\vartheta},\;\;\;\;\;\;
{\bf{ A}}_{1s}(p)=-{\bf e}_{1s}\sqrt{\hbar\omega}({\cal G}_1{\cal G}_2)^{1/2}\sqrt{\Delta\vartheta}\left( \frac{g_{10}}{g_{01}}\beta\right)^{\frac{1}{2}}.
\end{eqnarray}

Corresponding intensities of the monochromatic $\gamma$-radiation  are determined by
\begin{equation}\label{}
I_{\nu s}(p,\omega)=|\Psi^{(\nu)}_{\omega s}({\bf r})|^2.
\end{equation}

After introduction of the Authier's \cite{Authier} notation
\begin{equation}\label{}
\frac{1}{\Lambda_L}=\frac{1}{\tau_L}+i\frac{1}{\sigma_L},
\end{equation}
we are led to the following intensity distribution through the basis of the Borrmann triangle $(|p|<1)$ for the refracted beam:
\begin{eqnarray}\label{eq:F11}
I_{0 s}(p,\omega)=
\frac{|{\bf A}_{0s}(p)|^2}{\sqrt{1-p^2}}\left(\frac{1+p}{1-p}\right)e^{-\mu D}
\frac{2|\Lambda_L|}{D}\\ \times
 \left[\sinh^2\left(\frac{\pi D}{\sigma_L}\sqrt{1-p^2}+b \right)+\cos^2\left(\frac{ \pi D}{\tau_L}\sqrt{1-p^2}+\frac{\phi_2-\phi_1}{2}+\frac{\pi}{4}
 \right) \right],\nonumber
 \end{eqnarray}
and for the diffracted beam:
\begin{eqnarray}\label{eq:F12}
I_{1 s}(p,\omega)=
\frac{|{\bf A}_{1s}(p)|^2}{\sqrt{1-p^2}}e^{-\mu D}
\frac{2|\Lambda_L|}{D}\\ \times
\left[\sinh^2\left(\frac{\pi D}{\sigma_L}\sqrt{1-p^2}+b \right)+\sin^2\left(\frac{ \pi D}{\tau_L}\sqrt{1-p^2}+\frac{\phi_2-\phi_1}{2}+\frac{\pi}{4}
 \right) \right],\nonumber
\end{eqnarray}
where
\begin{equation}\label{}
\mu=\frac{1}{2}\left[\frac{\mu_0}{\gamma_0}+\frac{\mu_1}{\gamma_1}+
p\left(\frac{\mu_0}{\gamma_0}-\frac{\mu_1}{\gamma_1}\right)\right]
\end{equation}
with
\begin{equation}\label{}
\mu_{\nu}=\kappa\mbox{Im}g_{\nu\nu}=\sigma_a({\boldsymbol \kappa}_{\nu})/v_0,
\end{equation}
and $\sigma_a({\boldsymbol \kappa}_{\nu})$ meaning the absorption cross sections by an elementary cell
  of $\gamma$-quanta, which incident slightly off Bragg position with   the wave vectors $\approx{\boldsymbol \kappa}_{\nu}$. In accordance with the optical theorem \cite{Sitenko} the absorption cross section is determined by the imaginary part of the coherent elastic scattering amplitude to zeroth angle:
\begin{equation}\label{222}
\sigma_a({\boldsymbol \kappa}_{\nu})=\frac{4\pi}{\kappa}\mbox{Im}F({\boldsymbol \kappa}_{\nu},{\boldsymbol \kappa}_{\nu}).
\end{equation}
\begin{figure}[h]
\includegraphics[width=7cm]{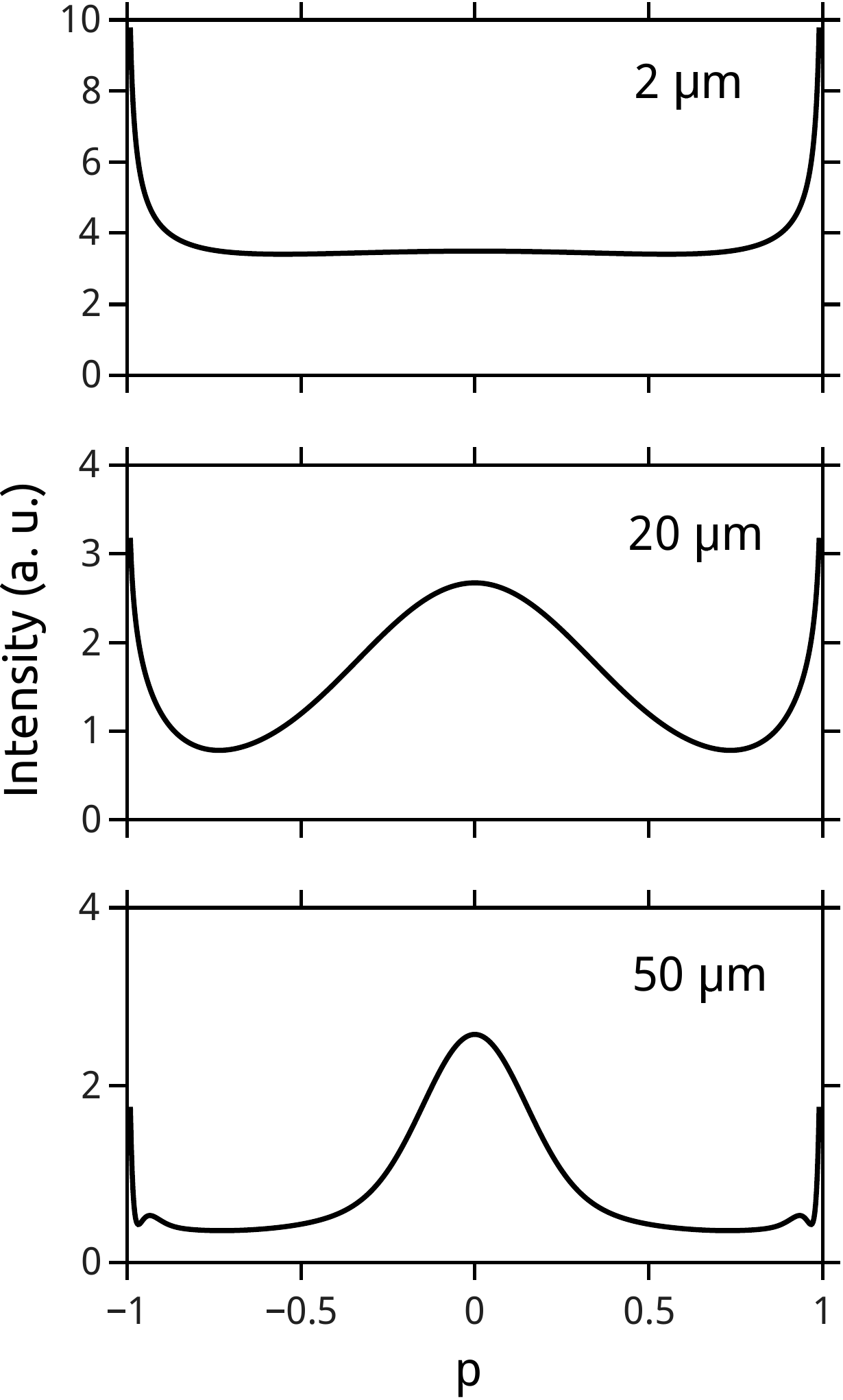}
\caption{Distribution of the M\"{o}ssbauer beam intensity  $I_1(p)$, diffracted in the tin film of different thickness, through the basis of the  Borrmann triangle for the symmetric  Laue diffraction  at the Bragg angle $\theta_{\textrm{\scriptsize
{B}}}=5^06'$, exact resonance $\omega_0=\omega'_0$, and the 100\% abundance  by the M\"{o}ssbauer isotope $^{119}$Sn. }
\end{figure}

\begin{figure}[h]
\includegraphics[width=7.5cm]{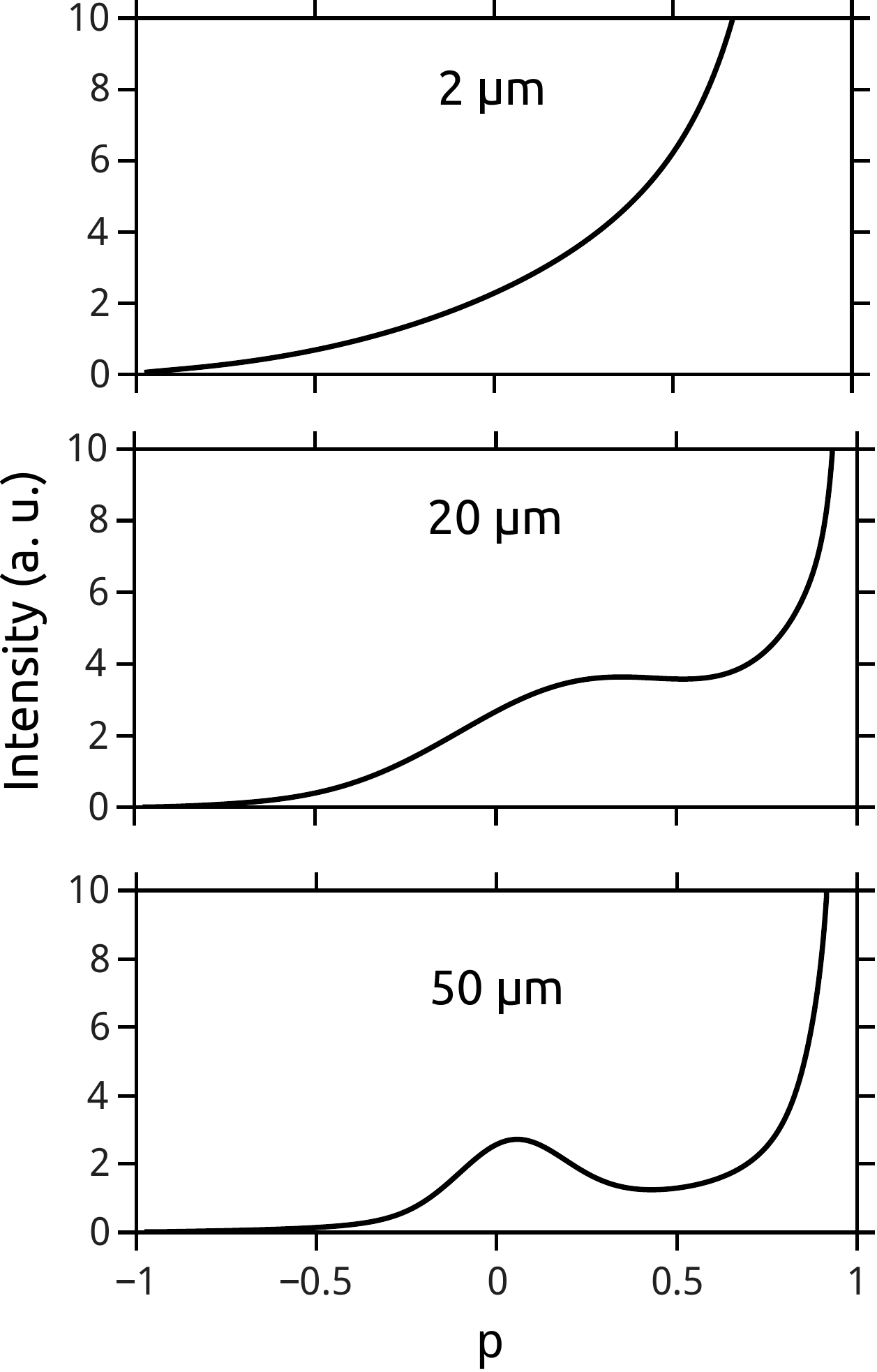}
\caption{The same as in Fig.2 but for the refracted beam $I_0(p)$. }
\end{figure}

In the opposite case of extremely narrow angular distribution, when
\begin{equation}\label{}
\sigma<<2\sqrt{\frac{|\Lambda_L|}{\pi D}  }|\Delta\vartheta|,
\end{equation}
 the angular distribution (\ref{eq:G}) behaves like a delta function,
\begin{equation}\label{}
{\cal G}_a(\eta)\approx \delta(\eta-\bar{\eta}).
\end{equation}
As a result, the integral in Eq.~(\ref{eq:10}) is estimated as
\begin{equation}\label{}
{\cal I}_{\nu s}^{(\iota)}=C_{\nu s}^{(\iota)}(\bar{\eta}) \exp\{{\cal N}_s
{\cal S}_{\iota s}(\bar{\eta})\}.
\end{equation}

At last, in the intermediate case of
\begin{equation}\label{}
\sigma \sim2\sqrt{\frac{|\Lambda_L|}{\pi D}  }|\Delta\vartheta|,
\end{equation}
the saddle points are determined by the equation
\begin{equation}\label{}
\frac{\eta-\bar{\eta}}{2\overline{\eta^2}}+{\cal N}{\cal S}'(\eta)=0,
\end{equation}
which reduces to an algebraic equation of the fourth order, giving already four saddle points.

It can be realized in experiments with synchrotron rays, when the angular dispersion achieves values
 $\sigma \sim 0.1|\Delta\vartheta|$  \cite{Authier}.

 Experimentally measured intensities of the $\gamma$-beams are obtained from (62), (63) by averaging them with the weight $|G_e(\omega)|^2$:
 \begin{equation}\label{}
 I_{\nu s}(p)=\int_0^{\infty}d\omega |G_e(\omega)|^2 I_{\nu s}(p; \omega).
 \end{equation}

\section{Discussion}
We developed  general dynamical theory   for the Laue  diffraction of divergent beams of $\gamma$-quanta, taking into consideration both their scattering
by atomic electrons and nuclei with low-lying excited levels. We confined ourselves by analysis of the two-wave  case, allowing the analytical solution.
The derived equations describe the refracted and diffracted beams for arbitrary orientation of the incident beam with respect to the Bragg resonance, i.e., for arbitrary deflection angle $\theta'_{\textrm{\scriptsize {B}}}-\theta_0.$  Therefore they may be useful for interpretation of experiments taking the rocking curves, when the target rotates with respect to the incident beam. In the case when $\theta'_{\textrm{\scriptsize {B}}}=\theta_0$
and respectively $\phi_1=\phi_2$ as well as $b=0$,
our equations (\ref{eq:F1})-- (\ref{eq:F12}) formally coincide with those given in the book \cite{Authier}. However, our formulas depend on the scattering amplitudes on nuclei absent in the formalism of \cite{Authier}. Moreover, they contain the realistic angular distribution of incident $\gamma$-photons.
It is approximated by the Gaussian function with arbitrary width $\sigma$,
which can be of the same order or even less compared to the diffraction interval $|\Delta\vartheta|$.
In principle, our final formulas remain the same if this Gaussian is replaced by any other function $G_a(\theta)$ normalized  by
\begin{equation}\label{}
\int_{-\infty}^{\infty}G_a^2(\theta)d\theta=1.
\end{equation}
Note also that according to derived formulae (65), (66) the deviation of the incident beam orientation $\theta_0$ from the Bragg angle $\theta_B$ leads to a shift of the Pendell\"{o}sung fringe.

\begin{figure}[h]
\includegraphics[width=7cm]{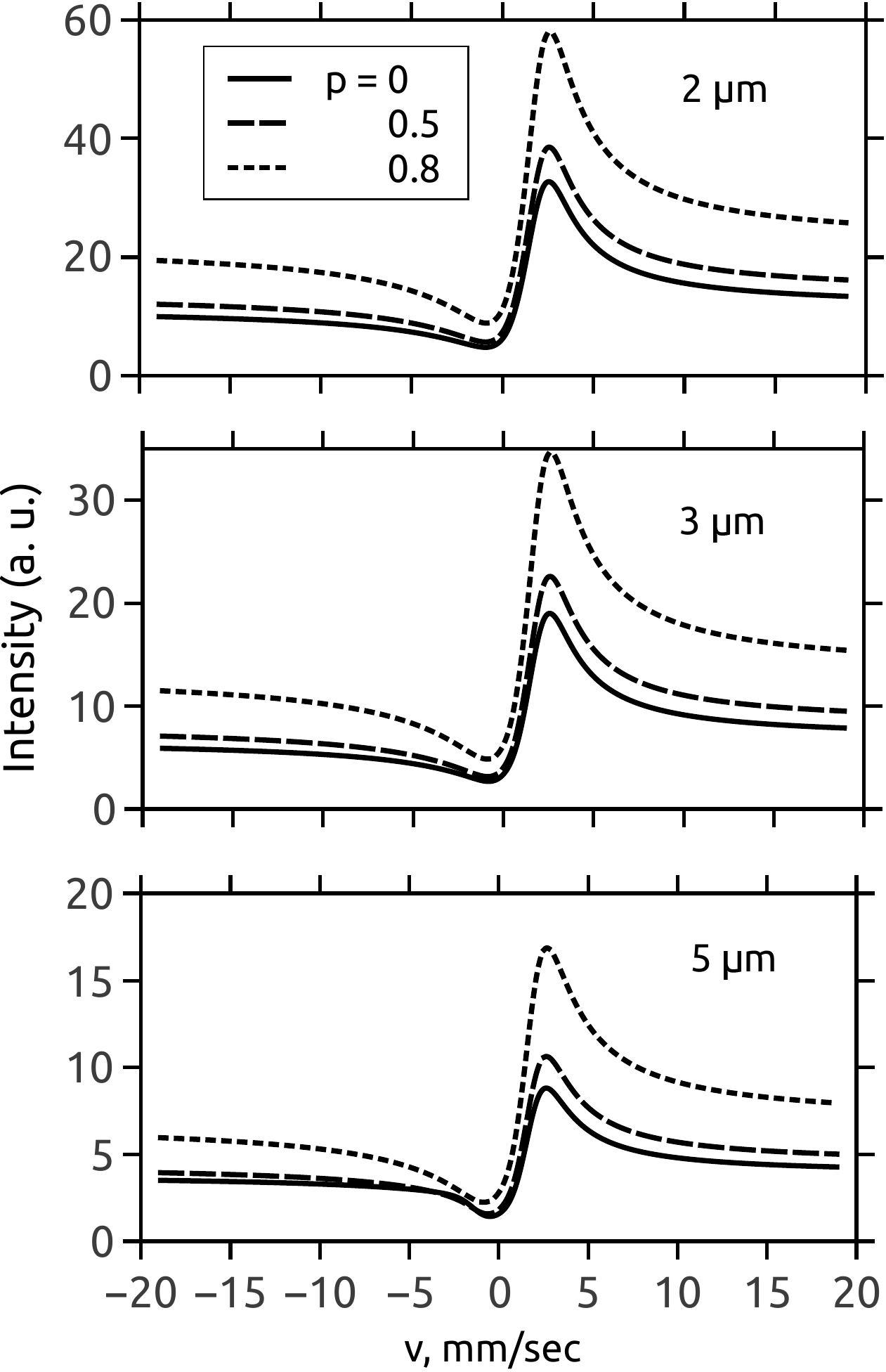}
\caption{The M\"{o}ssbauer spectrum of the diffracted beam  versus  the relative motion velocity of the  M\"{o}ssbauer source and the target for different positions of the scanning slit $p$. }
\end{figure}

\begin{figure}[h]
\includegraphics[width=7cm]{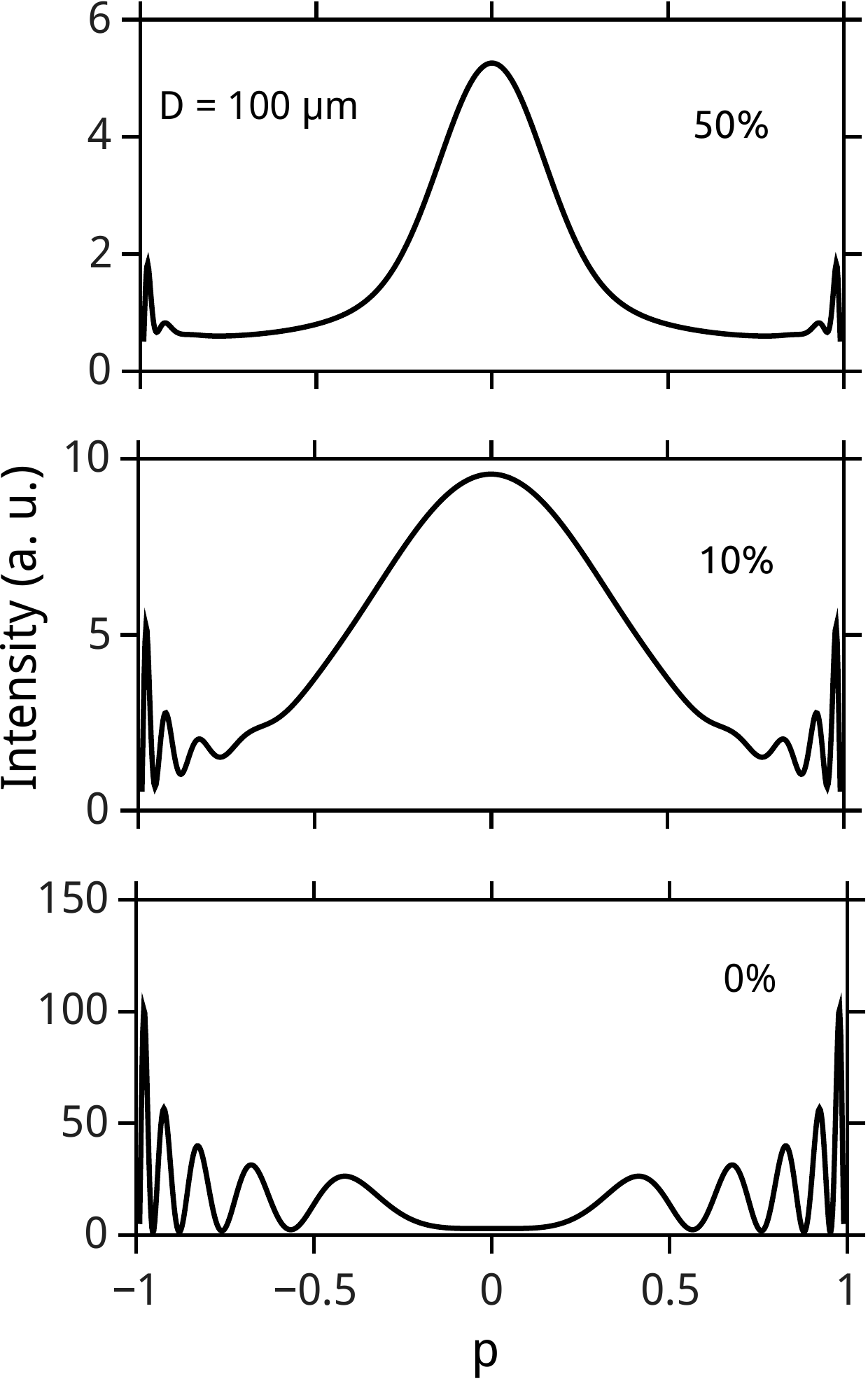}
\caption{Dependence of the intensity of diffracted radiation $I_1(p)$ on $p$ for  concentrations of the M\"{o}ssbauer isotope $^{119}$Sn $c_0=0.5,\;0.1$ and 0. }
\end{figure}

We illustrated our theory by analysis of the M\"{o}ssbauer diffraction, assuming
 the nuclear sublevels to be unsplit,  that simplifies consideration of the polarizations of $\gamma$-quanta.
 In particular, this is realized in the nuclei $^{119}$Sn with M1 transitions and  transition energy
$E_0=23.8$ keV. The M\"{o}ssbauer diffraction in tin single crystal has been observed by Voitovetskii et al. \cite{Voit1}, studying the suppression
of inelastic channels and reactions \cite{Kagan}.
We performed numerical calculations for the symmetric Laue diffraction in the tin crystal film (see Figs.~2-5), choosing the same parameters as in  the experiments reported in \cite{Voit1}. Namely, we  consider the first-order reflection by the  (020) planes with the Bragg angle $\theta_B= 5^06'$ and put the temperature $T=110$ K, when
   the  ratio of the resonant nuclear amplitude $|f^N_{res}|$ to the Rayleigh one $f^R$ equals 3.2 and the ratio of the absorption coefficients $\mu_N/\mu_e= 167$ \cite{Voit1}.  Taking into account that  the Debye temperature of tin $\Theta_D=200$ K we found the Lamb--M\"{o}ssbauer factor to be $e^{-W}=0.48$ at $T=110$ K, which enabled us to get the nuclear scattering amplitude.  
    All the  curves are calculated in the Kato's approximation   $\sigma>>|\Delta\vartheta|$. 
   All the curves drawn  in Figs.~2--4 are calculated for the 100\% abundance by  $^{119}$Sn. The results presented in Figs.~2,\;3 and 5 correspond to exact nuclear  resonance, when $\omega_0=\omega'_0$.

The calculated  intensities of the diffracted  beam $I_1(p)$ are shown in  Fig.~2
as a function of the parameter $p$, ranging from -1 to +1  for the film thicknesses $D=2,\; 20$ and  $50\;\mu$m. The corresponding curves for the refracted beam are given in Fig.~3.
Here we observe the same behavior as in the case of x-ray diffraction. Namely, in thin weakly absorbing crystal there is a growth of the diffracted intensity to the edges of the Borrmann triangle $p=\pm 1$.  With increasing film thickness, as the absorption grows, there appears a bump in the middle of the triangle  $(p=0)$. It can be explained that in strongly absorbing crystal the energy of $\gamma$-rays flows  mainly along the reflecting planes (see also \cite{Batterman}).

Dependence of the diffracted beam intensity on the relative velocity of  the M\"{o}ssbauer emitter and the target is analyzed in Fig.~4, where
   the crystal  thickness is taken $D=2,\; 3$ and  $5\;\mu$m. The intensity curves  manifest a characteristic asymmetry, caused by  interference of the waves coherently scattered by the  atomic electrons and the waves scattered by the nuclei. These results well agree with the observations of Voitovetskii et al. \cite{Voit1}.

The  diffracted wave intensity $I_1(p)$ versus the concentration of the M\"{o}ssbauer isotope $c_0$ is
   displayed in Fig.~5  for the same tin crystal but with thickness $D=100~\mu$m and  concentrations $c_0= 0.5,\;0.1$ and 0. We see that with lowering $c_0$ there appears fringe structure of the curve $I_1(p)$, which becomes most clear in the case  $c_0=0$, corresponding to pure Rayleigh scattering of  the M\"{o}ssbauer radiation.  With growing $c_0$ the oscillations of $I_1(p)$ are only conserved  at the edges of the Borrmann triangle.
 For $c_0=1$  oscillations of the curves $I_{1}(p)$ are absent.

Thus, in the Rayleigh scattering of M\"{o}ssbauer radiation experiments  one can observe the intensity oscillations of the diffracted waves. Numerical calculations show the introduction of the angular distribution with  $\sigma\sim |\Delta\vartheta|$  leads to significant distortion of these curves. Specifically, the diffraction curve  collapses to single peak at $p=0$ when the ratio $\sigma/|\Delta\vartheta|$ tends to zero.

Our theory may be useful in the structure analysis of crystals and especially in studies of crystal defects analogous to  \cite{Khrupa}.
Application of the M\"{o}ssbauer spectroscopy in such studies has great advantage compared to standard x-ray optics due to extremely narrow frequency distribution of M\"{o}ssbauer radiation.


\begin{thebibliography}{00}
\bibitem{Zach} W.H.Zachariasen, {\it Theory of X-ray Diffraction in Crystals} (Wiley, New York, 1945).
\bibitem{Batterman}B.W.Batterman,  H.Cole,  Dynamical diffraction of X Rays by
perfect crystals, Rev. Mod. Phys. {\bf 36}, 681 (1964).
\bibitem{Authier} Andr\'{e} Authier, {\it Dynamical Theory of X-ray
Diffraction} (Oxford University Press Inc., New York, 2001).
\bibitem{Kagan} A.M.Afanas'ev,  Yu.Kagan, Suppression of inelastic channels in resonant nuclear scattering in crystals,
 Sov. Phys. JETP  {\bf 21}, 215 (1965).
 \bibitem{Dz} A.Ya.Dzyublik,  Effect of forced vibrations on  scattering  of
X-Rays and M\"{o}ssbauer radiation by a crystal, phys. stat. sol. (b) {\bf 123}, 53 (1984);  {\bf 134}, 503 (1986).
\bibitem{Kato1} N.Kato, The energy flow of X-rays in an ideally perfect crystal:
comparison between theory and experiments,
Acta Cryst.  {\bf 13}, 349 (1960).
\bibitem{DS} A.Ya.Dzyublik, V.Yu.Spivak,
Laue diffraction of spherical M\"{o}ssbauer  waves, Ukr. J. Phys.  {\bf 61}, 826 (2016).
\bibitem{DSM} A.Ya.Dzyublik, V.I.Slisenko,  V.V.Mykhaylovskyy, Symmetric Laue diffraction of spherical neutron waves in
absorbing crystals,
 Ukr. J. Phys. {\bf 63}, 174 (2018).
\bibitem{DSM1} A.Ya.Dzyublik, V.V.Mykhaylovskyy, V.Yu.Spivak, Peculiarities of Laue diffraction of neutrons in
strongly absorbing crystals, Zh. Eksp. Teor. Fiz. {\bf 155}, 413 (2019); arXiv:1902.02051v1 [cond-mat.mtrl-sci]; JETP {\bf 128},  355 (2019).
 \bibitem{P1} J.Bialynicki-Birula, On the wave function of the photon, Acta Phys. Polonica {\bf 86}, 97 (1994)
\bibitem{Goldberger}  M.L.Goldberger, K.M.Watson, {\it Collision Theory}, (J.  Wiley,
New York, 1964).
\bibitem{Sitenko} A.G.Sitenko, {\it Scattering Theory} (Vishcha Shkola, Kiev, 1975; Shpringer, Berlin Heidelberg, 1999).
\bibitem{Belyakov} V.A.Belyakov,   Diffraction of M\"{o}ssbauer gamma rays in crystals,
Soviet Phys. Uspekhi {\bf 18}, 267 (1975).
 \bibitem{Math} M.A.Lavrentiev, B.V.Shabat, {\it Methods of the Theory of Functions of Complex Variable}
 (Nauka, Moscow, 1973).
\bibitem{Voit1} V.K.Voitovetskii et al., Diffraction of resonance $\gamma$ rays by nuclei and electrons in tin crystals,
Soviet Phys. JETP {\bf 27}, 729 (1968).
\bibitem{Khrupa} V.I.Khrupa, D.O.Grigoryev, A.Ya.Dzyublik, Structure perfection study of crystals containing micro- and macrodistortions
by x-ray acoustic method, Acta Phys. Polonica A  {\bf 86}, 597 (1994).
\end{thebibliography}
\end{document}